# Development of a multi-timescale method for classifying hybrid energy storage systems in grid applications

1st Christina Zugschwert
*Technology Center Energy*
*University of Applied Sciences Landshut*
Ruhstorf an der Rott, Germany
christina.zugschwert@haw-landshut.de

2nd Sebastian Göschl
*Technology Center Energy*
*University of Applied Sciences Landshut*
Ruhstorf an der Rott, Germany
sebastian.goeschl@haw-landshut.de

3rd Federico Martin Ibanez
*Center for Energy Science and Technology*
*Skolkovo Institute of Science and Technology*
Moscow, Russia
FM.Ibanez@skoltech.ru

4th Karl-Heinz Pettinger
*Technology Center Energy University of Applied Sciences Landshut*
Ruhstorf an der Rott, Germany
karl-heinz.pettinger@haw-landshut.de

*Abstract*— **An extended use of renewable energies and a trend towards increasing energy consumption lead to challenges such as temporal and spatial decoupling of energy generation and consumption. This work evaluates the possible applications and advantages of hybrid energy storage systems compared to conventional, single energy storage applications. In a mathematical approach, evaluation criteria such as frequency, probability of power transients, as well as absolute power peaks are combined to identify suitable thresholds for energy management systems on a multi-timescale basis. With experimental load profiles from a municipal application, an airport, and an industrial application, four categories, clustering similar roles of the VRFB and the SC, are developed.**

## I. Introduction

As the electricity demand in Europe is covered by an increasing share of renewable energy sources (RES), more impacts on stability, reliability and power quality occur in the electrical grid. These issues are mainly caused by the intermittency and decentralized placement of renewable power generation units.

In order to separate consumption and generation in terms of time and space, the use of energy storage systems (ESS) is already increasing. Currently, single storage applications based on lithium-ion batteries (LIB) or redox flow batteries (RFB) are used to reduce the impact of the difference between consumption and generation time-discretely. Thus, single energy storage technologies are dedicated to operate in defined time scales, often classified in the categories: milliseconds, seconds, up to hours, up to days [1].

The strategy to tailor storage systems for special loads, generators, and time scale requirements is an important part to reduce the dependency on fossil fuels and emissions as well as to provide future ESS for diverse applications [2]. Furthermore, the ability to create a storage solution operating on all timescales, from seconds to days, can be beneficial [3].

Hybrid energy storage system (HESS) operate on multiple time scales, allowing multiple services to be provided simultaneously, while high power peaks and high energy demands can be handled. Unfortunately, there is no single ESS yet on the market that has both features alongside with an affordable price. A vanadium redox flow battery (VRFB) has the capability to scale the energy density independently from the power setup. Although, the electrochemical reactions in a VRFB are not time limiting and some authors have reported battery's reactions times below 1s [4]. Typical power transients could take more time than these ranges due to startup procedures.

A supercapacitor (SC) has much higher power density and faster reaction time compared to VRFB [5, 6]. The expansion of the capacity entails an enormous cost for this system. Combining both technologies enables the HESS to act quickly, absorb high power peaks and additionally provide an extended service life at acceptable costs.

Currently, there are other HESS that combine LIB and SCs [7]. This is because LIB are often cheaper and more established in the market [8]. Replacing the LIB with a VRFB has certain advantages in stationary applications [8], especially if the physical system size is not a decisive factor. Advantages of a VRFB compared to a LIB are, safety, non-flammability, utilization of non-critical raw materials like Vanadium instead of Nickel or Cobalt, as well as an independent scalability of power and energy.

Within the scope of the EU-funded research project HyFlow[1], a high-power VRFB combined with SCs, operating as two distinct systems are electrically hybridized to optimize power and energy requirements. A lab scale demonstrator (5 kW, 10 kWh) as shown in **Fehler! Verweisquelle konnte nicht gefunden werden.** will be built during the project to verify the components and to develop a tailored energy management system (EMS).

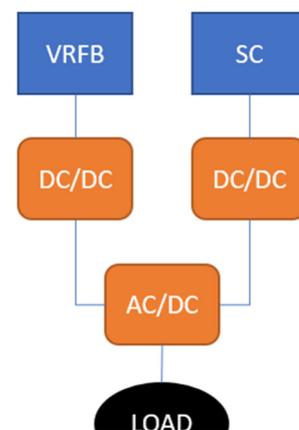

*Figure 1: Structure of the HyFlow HESS*

[1] HyFlow has received funding by the Horizon2020 research and innovation program of the European Union under grant agreement No 963550. https://hyflow-h2020.eu/.

The time dependent electricity demand of each consumer in the grid is represented by load profiles. They can be divided into a base load and a peak load. The latter arises from a short-term high demand for electrical energy or from the elimination of conventional power plants [9]. Studies on load profile data management assess maximum, minimum and average values to evaluate load management scenarios. Moreover, the load factor (relation from average power to maximum power) can be calculated and serves as an evaluation criterion [10]. Other approaches classify load profiles by percentiles [11] or apply mostly statistical models to analyze load values [12]. Furthermore, attempts are made to find recurring patterns in seasonal or weekly routines, depending on the specific load profile.

The main objective of the presented work is to develop an open source database for load profiles as well as to evaluate simple load profiles data management approaches by combined evaluation parameters. Universal mathematical approaches should lead to a better understanding of the classification of load scenarios for HESS applications. Although the operating strategy can be formulated in general terms to a large extent, the adaption for each load profile to achieve ideal results is beyond the scope of this work.

The paper is structured as follows: Section II describes the possible application categories of the HESS. Section III presents the experimental data and explains the algorithms used for the data analysis. Section IV shows the mathematical approach for a generalized evaluation process. Section V shows the results of the procedure. Section VI gives an outlook and concludes the paper.

## II. APPLICATION CATEGORIES FOR HESS

The pivotal applications for a HESS lie in the individual layout of each power and energy component and therefore adding more functionality or better cost advantages over the layout of a single component. The application cases shown in this paper require a storage solution, which can operate on various periods – short times as well as long durations.

The scope of the project is to set up an open access database including all collected load profiles and identify universal evaluation parameters for application specific control strategies for HESS. Table I shows the four basic application categories as well as the current data availability. [13].

The four categories can be divided into specific sub-categories clustering similar roles of the VRFB and the SC. One important application for peak shaving is the non-planned recuperating or transient processes in industry grids (e.g. engine test bench, sawmill, elevators, ships, etc.). Peaks occur for a short period, show high transients and could be in a non-planned schedule. There, a HESS has advantages in comparison to a single power or energy component. Some of these sub-applications e.g. elevators, or handling racks already use a SC for short power peaks. An additional VRFB enables an even higher utilization of renewable energy sources due to longer energy storage impact.

*Table 1: Application categories for experimental load data within the project HyFlow [13]*

| Category | Acronym | Data availability |
|---|---|---|
| Peak shaving of short and mid duration power peaks | PS | 4 data sets, 1-10 second resolution |
| Balancing of renewable energy generation in weak distribution grids | WDG | 2 data sets, 1 second resolution |
| Uninterruptible power supply | UPS | 4 data sets, 1 second resolution |
| Virtual inertia (momentary reserve) | VI | No data available yet |

Industrial grids, as a second sub-category, often show high-power peaks only on a few days of the year. Thus, SCs can cover these peaks, while the VRFB is getting ready to work. This operation mode is valuable because it reduces the high-power pulses and the related energy costs. Typically, the highest-power pulses per year is used to calculate the costs for industry grids. The higher utilization of the VRFB is related to a potentially higher utilization of RES that could lead to energy and financial savings.

Similarly, hybridization has advantages allowing fast charging applications and could improve weak distribution grids by balancing local active and reactive peak demands while shifting mid-day RES peaks to the evening [14]. The VRFB can work on a daily basis and is doing both power and energy services for longer durations depending on the penetration of RES. SCs are only applicable for high power peaks or high transients which means it has to work on a daily or weekly basis.

Moreover, the proposed HESS could be a more environmentally friendly and more flexible solution for UPS applications compared to the state of the art e.g. lead acid batteries or diesel generators [15]. The last category, virtual inertia (VI), is a solution for a future problem, in which conventional power plants are no longer available to a certain extent. Additional grid services, like controlling and operating micro grids or island grids - as shown in [14, 15], could also be adopted to the presented HESS.

Besides the scope of that work, a HESS could utilize braking energy e.g. in subway stations as shown in [17]. In general, hybridization of VRFB and SC adds dynamic to the system and increases operability in terms of efficiency, cycle life time of components and potentially costs [13].

## III. EXPERIMENTAL PART

Up to now, seven load profiles are assigned to the PS category. An airport power profile was extracted during January 2021 (one month) in a time resolution of 15 minutes. An industry grid, on the other hand, covers the whole year of 2019, with the same time resolution. Unfortunately, 15 minutes time resolution is not accurate enough for a statement about power distribution between the HESS components as the SC is considered to work in a time frame from milliseconds up to 30 seconds [13].

Figure 2 shows the difference between two different time scale profiles. The red profile is measured within one second time step, whereas the black profile is recorded in 15-minute intervals. Steep peaks with high load frequency are visible at the one second level, while they are balanced out with average values at the 15-minute level. The black load profile runs much smoother. The low time resolution of 15-minutes is applicable for a battery-only storage application, as the main purpose of an EES is to balance load and generation over time periods of several hours. SCs are used to cover high and steep load peaks in a short period of time. Component management systems need load profiles to control the SC with a similar high time resolution. Thus, all load profiles with less time resolution than 10 seconds are not suitable for further use.

Furthermore, the use of profiles with higher time resolution (at least 10 seconds) is recommended for reasons of realism. In order to develop control algorithms for real operation, the HESS must be able to react to load fluctuations on all time-scales.

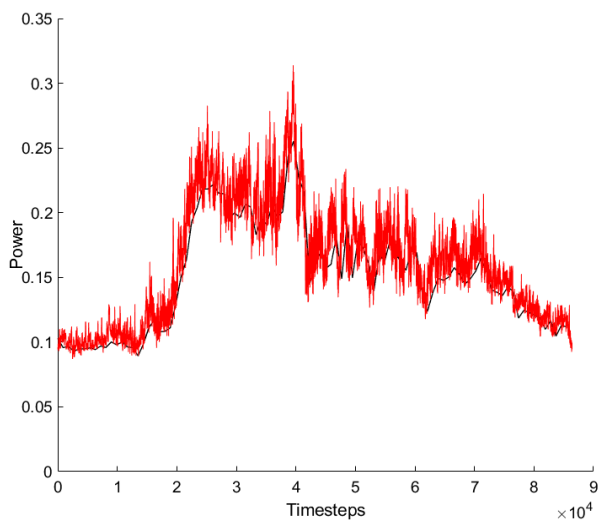

*Figure 2: Comparison between 1s and 15min Load Profile*

In order to provide high time resolution load profiles of one second within the scope of this paper, two research institutes located in Germany and Russia recorded their load profiles. Therefore, data from both University of Applied Sciences Landshut (Germany) and the Skolkovo Institute of Science and Technology (Russia) are measured. Additionally, a synthetic load profile of an industrial plant with a 10 second resolution was considered.

Furthermore, load profiles from a municipality in lower Bavaria has been measured for WDG category. The dataset is available in one second resolution and shows the power supply of a whole city district with a mixed consumer structure. Both family homes and small production sites are located in this area.

A second dataset from a municipality in Bosnian Herzegovina covers three different areas with mixed generation and consumer structures with a time resolution of 15 minutes. As already explained above, the data set has not been selected for further use, as the time resolution is not suitable. Every profile with a time resolution below 30 seconds from the WDG and PS categories can be used for the third category (UPS). Therefore, the power requirement from the HESS is set to zero up to a certain point in time while the operation of the HESS is used to completely cover the required load. Category VI is not specified with load profiles yet. [11]

IV. EVALUATION APPROACH

The main goal of the design and operation of the HESS is to maximize the utilization of each storage component. This requires the appropriate control and dimensioning of both components, depending on the intended use and application case. The following sections deal with the mathematical evaluation approach to decide upon the load distribution between the individual components. In order to cover the widest possible range of applications, the physical dimensioning of the system is to be carried out at the latest possible stage. In the further course, the assumption is made that the complete load profile for the location of the storage system under consideration is already known or has been predicted by data science. During the investigation the data is normalized to the maximum peak power per load profile within the period under review. Thus, power values only vary between zeros and one.

The mathematical evaluation approach can be separated into three steps: (i) relative height of a load peak (ii) frequency of power transients, and (iii) calculation of power transients over time period. Afterwards, a combined interpretation leads to improved control strategies for HESS.

Considering the HESS as a black box, without classifying the components used, two factors can be applied to determine the system control: (i) the relative height of a load peak for dimensioning and design of the HESS and (ii) frequency of power transients (histogram). The HESS utilization is adapted by reviewing the frequency of power transients over time as the energy demand is represented discretely in time. Thus, classifications can be made for the different load ranges such as base or peak load. These areas can then be marked and analyzed in a time-continuous manner.

If we now look at the system not exclusively as a black box, but integrate the planned application purposes of both storage technologies - SC as a fast, but short option and VRFB as a mid-term, hourly storage technology - it becomes apparent that another decision factor is required: (iii) the time-derivative. The time-derivative of the load profile is suitable for this purpose because it describes frequency and duration of the power transients for each time step. It is a good index for understanding transients and will be used later on to decide how the load will be splitted between the SC and the VRFB. The higher the derivative, the more worthwhile is the use of the SC.

The challenge is to define an appropriate threshold of the power transient for the use of both the VRFB ($flag_{VRFB}$) and the SC ($flag_{SC}$). The threshold indicates at which relative energy demand the storage system is used either in addition to or instead of using the power grid. First, assume that a threshold of 80% of the maximum power demand and 80% of the highest discharge is suitable as a delineation of the application area of both storage technologies. A delimitation between the base load and the area of application of the VRFB is not possible across the board due to strongly varying base load levels.

$$flag_{SC} = \begin{bmatrix} 1 \; if \; P_{load}(t) > 80\% * P\_\max \\ 0 \; in \; any \; other \; case \end{bmatrix}$$

$$flag_{VRFB} = \begin{bmatrix} 1 \: if \: P_{load}(t) \leq 80\% * P\_max \\ 0 \: in \: any \: other \: case \end{bmatrix}$$

In the further course of the project, threshold values must also be found under which the individual components are to be reloaded. As an example the limit was set to 80% of the maximum power, but this threshold will depend on the application and further studies must be performed.

## V. RESULTS AND DISCUSSION

Particular use cases mentioned in Table 2 are analyzed according to the criteria mentioned above, as well as categorized for their suitability for HESS application. The load profile of a cutting machine of the Skolkovo Institute as well as an electrical vehicle (EV) charging park serve as an example for peak shaving applications (B and C). The distribution network of a municipal utility in Lower Bavaria, is analyzed in the WDG category (A).

*Table 2: Use cases within the scope of this paper*

| Category | Use Case | Sub-Category |
|---|---|---|
| Balancing of renewable energy generation in weak distribution grids | Municipal utility in Lower Bavaria | A |
| Peak shaving of short and mid duration power peaks | Cutting Machine | B |
| Peak shaving of short and mid duration power peaks | EV charging park | C |

### A. Distribution Grid in Lower Bavaria

Figure 3**Fehler! Verweisquelle konnte nicht gefunden werden.** shows the normalized load profile for sub-category A. The daily load curve is illustrated in the context of a twelve-day data set. Daily routines in the mixed residential area result in similar load patterns. The base load levels off at about 50% of the maximum load.

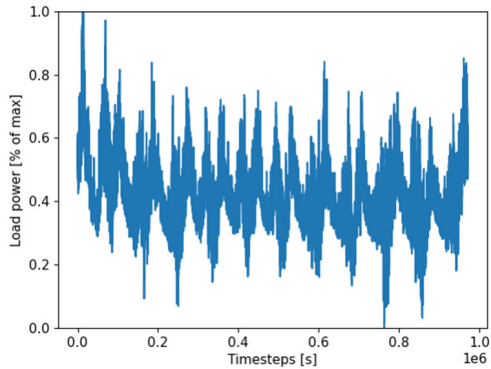

*Figure 3: Normalized load profile distribution grid in lower Bavaria*

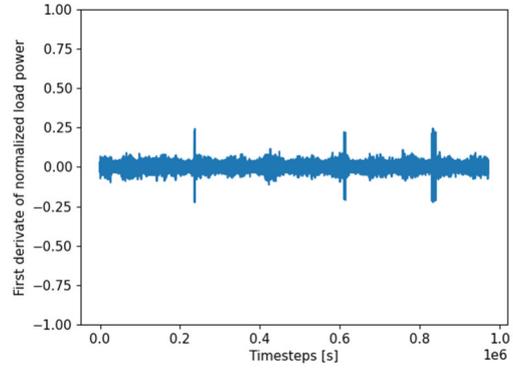

*Figure 4: Normalized time-derivative load profile distribution grid in lower Bavaria*

Figure 4 shows the normalized time-derivative of the load profile considered above. Except for a three higher peaks, only small peaks are visible. This shows that only very isolated high jumps in the load demand occur. This is explained by the combination of the load demand of many households, which balance out their demand over the course of the measurement.

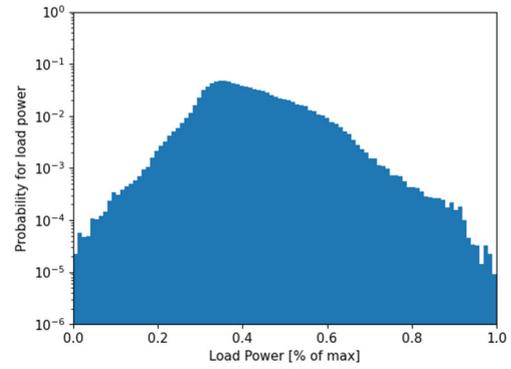

*Figure 5: Histogram of load profile distribution grid in lower Bavaria*

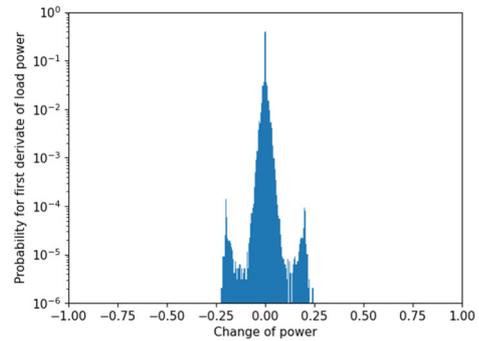

*Figure 6: Histogram time-derivate of load profile distribution grid in lower Bavaria*

Figure 5 shows the histogram of the load profile over the entire recording period. When looking at the graph, an approximately equal distribution is noticeable. The peak in the middle represents the power demand that was queried most often. This range coincides with the base load range mentioned above, so it can be assumed that the base load range defines up to a utilization of 40% of the maximum power. Furthermore, the range between 80% and 100% of the maximum load can now be defined as the possible operating range of the SC.

Figure 6 shows the histogram of the time-derivative over the entire period. Unlike the histogram of the load profile, this graph is axisymmetric to the y-axis and the zero point of the x-axis. Again, the assumption can be confirmed that the derivative stays in very low value ranges for most of the time, while another peak is formed at both the negative and positive maximum. These areas form the deployment times of the SC, since a rapid change in the power demand takes place there.

In summary, the load profile can be classified as HESS-compliant. However, peaks in larger grid areas are often balanced out by the large number of connected consumers. This leads to the conclusion that the level of investigation for application scenarios for HESS should be as small as possible. By focusing on a few households or consumers, which have different demand curves, the internal balancing possibility of the system is reduced and the SC, as one part of the HESS, is more applicable.

*B. Peak Shaving in Research Insitute*

Figure 7 shows the normalized load profile for sub-category B. Compared to the previous load profile, no continuous load demand takes place here. In fact, the machine is not used for half of the time during the period under consideration. Among other things, the utilization time is related to day-time, working days, and the associated working hours. A high energy demand can be determined if the machine is used. There are high switch-on peaks, while the energy requirement remains relatively constant after the switch-on process.

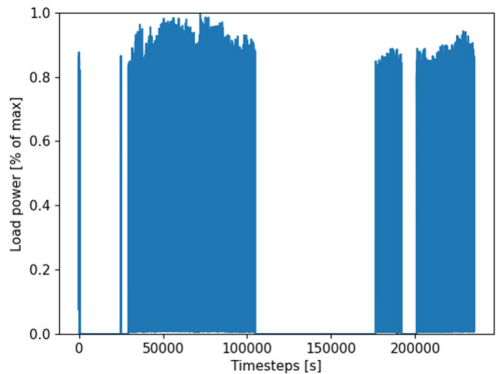

*Figure 7: Normalized Load Profile Research Institute*

The time-derivate of the load profile (Figure 8) shows significantly higher fluctuations compared to the previous deployment sub-category. As already stated, the time-derivate also shows that the cutting machine is not a constant but varying load. This is an ideal application for the SC.

Looking at Figure 9 and Figure 10, the hypothesis can be validated. The histogram of the charging power does not show a uniform distribution, but a collection of values around the zero point and the maximum. This means that either a lot of power or no power at all is required for most of the time.

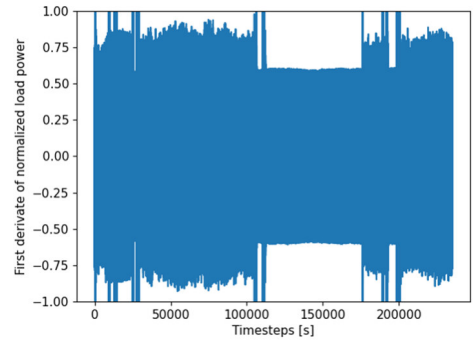

*Figure 8: Normalized Derivate Research Institute*

The histogram of the derivative also confirms this theory. Similar to the previous sub-category, the graph is approximately axisymmetric about the y-axis. However, in addition to the peak in the low derivative region, there are additional peaks in the high derivative region here. This favors the use of a HESS combining battery and SC.

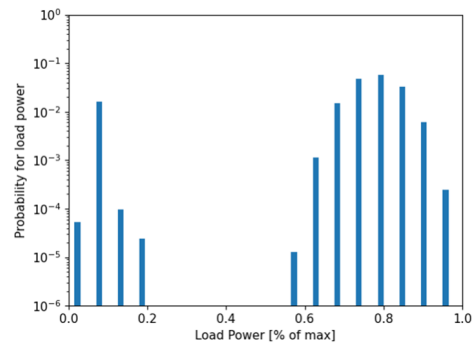

*Figure 9: Histogram of research institute*

Due to the small system size (only one machine) and complexity, the load demand is not compensated internally but passed directly to the HESS. This creates more application possibilities for the SC, both in terms of frequency of use and intensity of use.

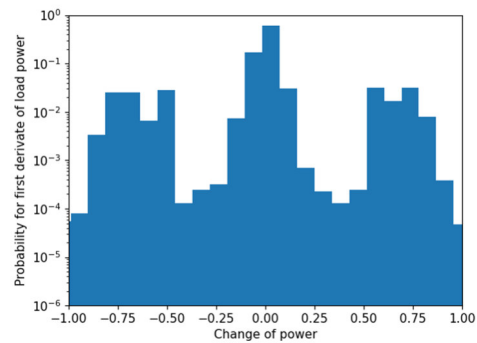

*Figure 10: Histogram of time-derivate Research Institute*

This use case shows the potential of the HESS: the isolated strong load peaks will be absorbed by the SC, while the VRFB buffers smaller deviations.

*C. Peak Shaving at EV charging park*

Sub-category C considers a peak load smoothing scenario at an electric vehicle (EV) charging park. The load profile

(Figure 11) shows a constant change in the required load over the entire period under consideration, which is caused by the vehicles arriving irregularly and unscheduled for charging.

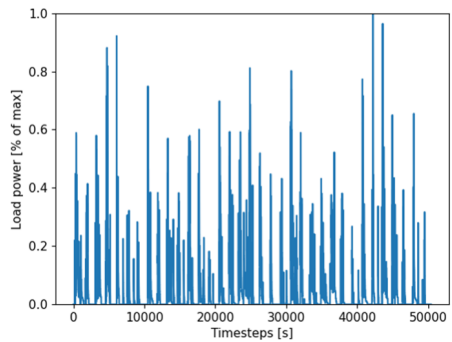

*Figure 11: Normalized Load Profile Charging Park*

In contrast to sub-category A, these unscheduled charging means that no regular sequence can be detected, as the charging processes can also overlap in time. It is assumed that the vehicles are charged independently of each other and therefore no smoothing is performed by the charging stations. The irregular utilization of the charging station results in both load peaks up to the maximum load and those that contain less energy demand. The base load can be set to zero in this sub-category.

The time-derivative (Figure 12) shows stronger positive peaks than negative ones. The vehicles immediately start with maximum charging power at the beginning of the charging process, while the required power is gradually reduced in the last time step of the charging process. Due to the abrupt start of charging and the associated high discharge, this may be an application with high relevance of the SC.

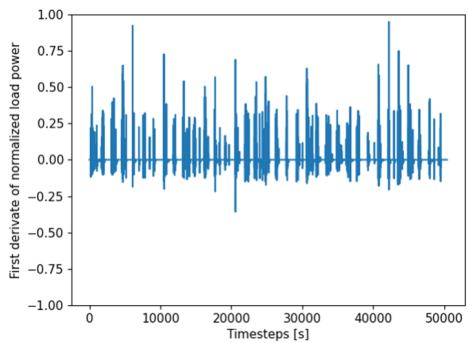

*Figure 12: Normalized Derivate Charging Park*

Compared to the two previous sub-categories, a difference can be seen when looking at the histogram of the time-derivative (Figure 13): the axis symmetry is not given in this example. Here, as in the continuous-time representation, it becomes clear that the negative derivative contains significantly lower values than the positive one. If we look exclusively at the range of the positive derivative, we again notice an implied uniform distribution with a peak at 0.1pu. Isolated peak loadings are also recognizable.

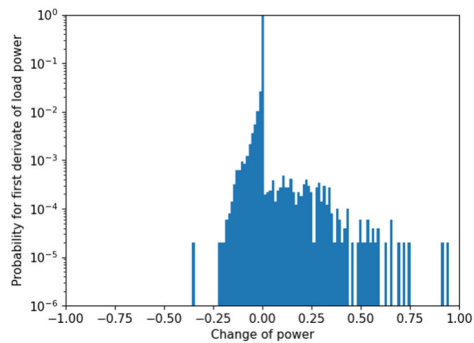

*Figure 13: Histogram of the Derivate Charging Park*

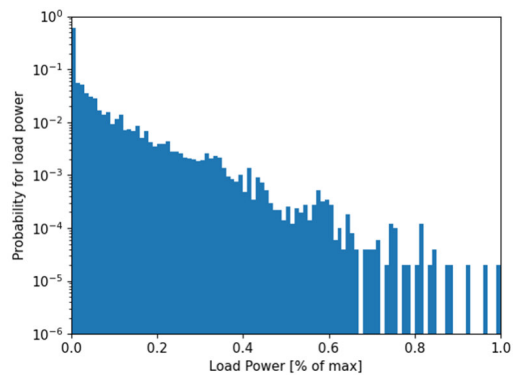

*Figure 14: Histogram of the Charging Park*

When looking at the histogram of the charging power, it is noticeable that it strongly resembles a linearly decreasing straight line. As can be seen in Figure 14, the load park rarely reaches maximum utilization, while low or no utilization occurs very frequently. However, a pure focus on the maximum utilization is not expedient here; the derivation must also be considered.

In summary, the charging park can be judged as a good scenario for a HESS. Both the use of a battery and the SC can be validated by the intensity and number of peak loads.

## VI. CONCLUSION AND OUTLOOK

Application scenarios of HESS are investigated based on real load profiles, subdivided into four application categories. Generally, the categories show different behaviors for the evaluation criteria. While both load profiles from research institutes (category PS) are mostly characterized by high fluctuations in the derivate and high average power, the load profile from the charging station shows high frequency of use, while the average power is mostly constant.

The data used, has been collected within the research project HyFlow. HyFlow develops a HESS with, among other, the aims of improved efficiency, flexible control algorithms, extended lifespan alongside with optimized costs and increased competitiveness to existing single storage technologies.

Using the approaches described in this paper, threshold values within the EMS will be used to predict all power demands precisely and dedicate the most efficient power flow between the VRFB and the SC. In most scenarios, the VRFB is used as a mid-term energy supply (minutes up to hours), while the SC will cover high power peaks (milliseconds up to seconds). The paper's goal of defining fixed thresholds universally for different application purposes turns out to be problematic. The specified decision factors must remain flexible in their values in order to cover as wide a range of applications as possible. For example, the vehicle charging park is an ideal deployment scenario, but a threshold value of 80% of the maximum would largely prevent deployment of the SC.

Continuous work will extend the approach by various predictive analysis methods and control strategies. In addition, the assumptions made so far will be confirmed by generating and analyzing new and more load profiles. The economic factor, which up to now has had no impact on the choice and combination of system sizes, will also be integrated into the calculations. Furthermore, the charging of the HESS must also be considered and implemented in the future work of the project. Especially in the case of deployment scenarios with a constant base load, an economic analysis is necessary in order to find the most cost- and energy-efficient deployment strategy.

ACKNOWLEDGEMENT

This research is funded by the Horizon2020 research and innovation program of the European Union under grant agreement No 963550, HyFlow. The authors thank the consortium formed for the HyFlow project and the external advisory board for contributing the experimental data and conceptual discussions.